\def\kB{k_{\text{B}}}

\def\be{\begin{equation}}
\def\ee{\end{equation}}
\def\bea{\begin{eqnarray}}
\def\eea{\end{eqnarray}}
\def\bse{\begin{subequations}}
\def\ese{\end{subequations}}

\documentclass[prl,twocolumn,showpacs,preprintnumbers,amsmath,amssymb]{revtex4}
\usepackage{graphicx}% Include figure files
\usepackage{dcolumn}% Align table columns on decimal point
\usepackage{bm}% bold math
\begin{document}
\preprint{}
\title{Quantum Phase Transition in a Clean Two-Dimensional Electron System}
\author{T.R. Kirkpatrick$^{1}$ and D. Belitz$^{2}$}
\affiliation{$^{1}$Institute for Physical Science and Technology, and Department
                   of Physics, University of Maryland, College Park, MD 20742\\
         $^{2}$Department of Physics and Theoretical Science Institute, University
                of Oregon, Eugene, OR 97403}
\date{\today}
\begin{abstract}
A quantum phase transition that was recently observed in a high-mobility silicon MOSFET 
is analyzed in terms of a scaling theory. The most striking characteristic of the transition
is a divergence of the thermopower, according to an inverse linear law, as a critical value
of the electron density is approached. A scaling description of this transition yields 
predictions about the critical behavior of other observables, e.g., the specific heat. We also 
explore the possibility that this transition realizes a recently predicted transition from a 
Fermi liquid to a non-Fermi-liquid state.
% 589 characters 
\end{abstract}
\maketitle
The question of the existence or otherwise of a metal-insulator transition in the bulk of a
two-dimensional ($2$-$d$) electron system has been an issue of great interest for some
time. It had seemed settled after it became clear that the lower critical dimension for the
Anderson metal-insulator transition of noninteracting electrons is $d_{\text{c}}^{\,-} = 2$ 
-- that is, the system is insulating and no transition to a normal-metal phase can take place 
in $d=2$ \cite{Abrahams_et_al_1979, Wegner_1979, AT_footnote} -- and generalizations of 
the pertinent theory concluded that this remains true for interacting electrons as well
\cite{Finkelstein_1984a, Belitz_Kirkpatrick_1994}. The observation of what appeared to be a 
metal-insulator transition in a strongly interacting $2$-$d$ electron system, namely, a 
silicon metal-oxide-semiconductor field-effect transistor (MOSFET) \cite{Kravchenko_et_al_1994} 
therefore came as a considerable surprise, as it indicated the possibility of an interaction-driven 
quantum phase transition that involves a normal metallic phase. However, despite a large amount 
of experimental and theoretical work, the physics underlying these observations, and even the 
existence of a true phase transition, continues to be debated 
\cite{Abrahams_Kravchenko_Sarachik_2001, Spivak_et_al_2010}. 

One problem is that the electrical conductivity, which is the most obvious observable to focus 
on in the context of a metal-insulator transition, is susceptible to complicated and non-universal
scattering mechanisms that can either mimic or mask true critical behavior. It is therefore very 
interesting that a recent experiment on low-disorder MOSFETs has found what appears to be 
critical behavior of the thermopower, or Seebeck coefficient, $S$ as a function of the electron 
density \cite{Mokashi_et_al_2012}. The thermopower is a ratio of two transport 
coefficients \cite{Kubo_formula_footnote}, and therefore presumably less susceptible to the 
non-critical scattering that can make transport coefficients hard to interpret. The experiment of 
Ref.\ \onlinecite{Mokashi_et_al_2012} found that $1/S$, at temperatures between $T = 300\,$mK 
and $800\,$mK, is a linear function of $1/T$ with a prefactor that vanishes approximately linearly 
as the electron density $n$ approaches a critical value $n_{\text{c}}$ from above:
\be
S(T,n) \equiv eT\,s(n) \propto T\,(n - n_{\text{c}})^{-\mu}
\label{eq:1}
\ee
with $e$ the electron charge, $\mu = 1.0 \pm 0.1$ and 
$n_{\text{c}} = (0.78 \pm 0.01)\times 10^{11}\,$cm$^{-2}$.
The sample in question also shows what Ref.\ \onlinecite{Mokashi_et_al_2012} interpreted
as a resistive transition at a slightly higher density, close to $n = 0.8\times 10^{10}\,$cm$^{-2}$. 
This feature is very similar to the putative metal-insulator transition in higher-disorder samples, 
and the density where it occurs is strongly disorder dependent, whereas the critical density 
$n_{\text{c}}$ for the thermopower is independent of the disorder. This led the authors of 
Ref.\ \onlinecite{Mokashi_et_al_2012} to suggest that their system, at the temperatures that were
accessible in their experiment, can be reasonably interpreted as a clean electron fluid with a normal
metallic phase. Under these assumptions, the critical behavior of $S/T$ signifies a transition
from a Fermi liquid to an unknown phase that is driven by electron-electron
interactions, not sensitive to weak disorder, and that is characteristic of, and would still be 
present in, a true clean system with no disorder. It thus joins the ranks of, but is distinct from,
other quantum phase transitions in the bulk of clean $2$-$d$ electron systems, e.g., the
antiferromagnetic quantum phase transition \cite{Abanov_Chubukov_2004}, or transitions 
involving stripe phases \cite{Kivelson_et_al_2003}. In what follows, we adopt this hypothesis.

In Ref.\ \onlinecite{Mokashi_et_al_2012} the observations summarized by Eq.\ (\ref{eq:1}) and
the above discussion were interpreted in terms of an
effective mass that diverges at $n_{\text{c}}$. We will come back to this interpretation below.
However, if they represent a true quantum phase transition, then it is natural to first perform 
a general scaling analysis \cite{Stanley_1971, Ma_1976}. To do so is the purpose of
the present Letter. We will initially give a very general analysis that relies on a minimal set of
assumptions, and then explore the possibility that the observed transition is related to a
recently proposed transition between a Fermi liquid and a non-Fermi-liquid in clean
electron systems \cite{Kirkpatrick_Belitz_2012}. Within the framework of this suggestion,
the order parameter for the transition is the density of states at the Fermi level, and the
low-density phase is a ``strange metal'' or "interaction-induced semimetal" characterized 
by a pseudogap in the density of states which represents the phase that 
Ref.\ \onlinecite{Mokashi_et_al_2012} postulated to precede a Wigner crystal.

Let us assume that the observations expressed by Eq.\ (\ref{eq:1}) reflect a quantum
phase transition with some unknown order parameter density ${\cal N}$ \cite{hidden_order_footnote}.
We assign scale dimensions $[L]=-1$, and $[t] = -z$ to factors of length and time, 
respectively. We use units such that $\hbar = \kB = 1$, so energy and temperature
both carry scale dimensions $[E] = [T] = z$. We denote the (unknown) field conjugate to
${\cal N}$ by $h$, with a scale dimension $[h] = y_h$.
Finally, let the dimensionless distance from the critical point be $t$ with a scale dimension 
$[t] = 1/\nu$, where $\nu$ is the correlation length exponent. For the experiment
in question, $t = (n - n_{\text{c}})/n_{\text{c}}$. $\nu$, $y_h$, and $z$ are in general
three independent critical exponents, and all other exponents can be expressed in terms
of these three \cite{Sachdev_1999}. Then the free energy density $f$, which 
dimensionally is an energy per volume, satisfies the homogeneity law
\be
f(t,T,h) = b^{-(d+z)}\,f(t\,b^{1/\nu}, T\,b^z,h\,b^{y_h})\ ,
\label{eq:2}
\ee
where $b$ is the length rescaling coefficient.
Now consider the thermopower $S$, which is defined as the ratio of the induced 
voltage gradient to an applied temperature gradient in the absence of an electrical 
current \cite{Kubo_formula_footnote}. $eS$ is thus dimensionless by power 
counting \cite{inequality_footnote}. At a phase transition where it displays critical
behavior it is expected to obey a homogeneity law
\bse
\label{eqs:3}
\be
eS(t,T) = eS(t\,b^{1/\nu}, T\,b^z)\ .
\label{eq:3a}
\ee
If we use the experimental observation that the thermopower is proportional to $T$,
and define $s = S/eT$, we have
\be
s(t) = b^z\,s(t\,b^{1/\nu}) \propto t^{-\nu z}\ .
\label{eq:3b}
\ee
\ese
Comparing with the results of Ref.\ \onlinecite{Mokashi_et_al_2012}, we can draw a first
nontrivial conclusion, namely, the correlation length exponent $\nu$ and the dynamical
exponent $z$ are related by \cite{multiple_z_footnote} 
\be
\nu z = 1\ .
\label{eq:4}
\ee

We next show how scaling leads to a prediction for the specific heat coefficient 
$\gamma \equiv C_V/T = -\partial^2 f/\partial T^2$. From Eq.\ (\ref{eq:1}) we find
\bea
\gamma(t,T) &=& b^{z-d}\,\gamma(t\,b^{1/\nu}, T\,b^z)
                      = t^{d\nu -1}\,f_{\gamma}(T/t^{\nu z})
\nonumber\\
                    &=& t^{2\nu -1}\,f_{\gamma}(T/t)\ .
\label{eq:5}
\eea
Here $f_{\gamma}$ is a scaling function, and in the second line we have specialized
to $d=2$ and used Eq.\ (\ref{eq:4}). This predicts that a scaling plot of $\gamma/t^x$
versus $T/t$, with $x$ suitably adjusted, will make data for various values of $T$ and $t$
collapse onto one curve \cite{energy_scales_footnote}. In particular, the specific heat
coefficient extrapolated to $T=0$ will vanish as $t^{2\nu - 1}$ as the transition is
approached from high densities. This will allow for an experimental 
determination of the correlation length exponent $\nu$, as well as the scaling function $f_{\gamma}$. 
From Eq.\ (\ref{eq:4}) one then obtains $z$.

A determination of $\nu$ will also provide a check for whether the transition can 
possibly be dominated by disorder -- which we assume it is not, see %Ref.\ \onlinecite{MIT_footnote}). 
the discussion after Eq.\ (\ref{eq:1}).
The Harris criterion \cite{Harris_1974, Chayes_et_al_1986} states that for a critical fixed 
point to be stable with respect to quenched disorder the inequality $\nu \geq 2/d$ must 
hold. While no conclusion can be drawn if $\nu$ is found to satisfy the Harris criterion,
a value $\nu < 1$ would rule out a disorder-dominated transition. 

Let us now turn to the unknown order parameter density ${\cal N} = -(\partial f/\partial h)/T$. 
From Eq.\ (\ref{eq:1}) we have
\bea
{\cal N}(t,T) &=& b^{\,y_h - d}\,{\cal N}(t\,b^{1/\nu}, T\,b^z) = t^{\nu(d-y_h)}\,f_{\cal N}(T/t^{\nu z})
\nonumber\\
          &=& t^{\nu(2-y_h)}\,f_{\cal N}(T/t)\ ,
\label{eq:6}
\eea
with $f_{\cal N}$ another scaling function.
If the nature of the order parameter were known, then a scaling plot in addition to the
one described by Eq.\ (\ref{eq:5}), combined with Eq. (4), would thus provide the 
third independent critical exponent at the quantum phase transition. All other
critical exponents can be expressed in terms of $\nu$, $z$, and 
$y_h$ \cite{Belitz_Kirkpatrick_1994}. For instance,
from Eq.\ (\ref{eq:6}) we see that the order-parameter exponent $\beta$, defined
by ${\cal N}(t,T=0) \propto t^{\beta}$, is given by $\beta = \nu(2-y_h)$, the exponents
$\gamma$ and $\eta$ that govern the dependence of the order-parameter
susceptibility on $t$ and the wave number, respectively, are given by
$\gamma = \nu (2-\eta) = \nu (2y_h - 2)$, etc.

We finally explore the possibility that the observed transition is the one from
a Fermi liquid to a non-Fermi liquid proposed recently in Ref.\ \onlinecite{Kirkpatrick_Belitz_2012}.
The order parameter for this transition is the density of states $N$ at the Fermi level, 
which dimensionally is an inverse energy times an inverse volume. If it is critical, it
is thus expected to have a scale dimension, in $d=2$, $[{\cal N}] = [N] = 2-z$, which implies 
\be
y_h = z\ .
\label{eq:7}
\ee
That is, at this transition there are only two independent critical exponents.
The scaling behavior of the density of states then becomes the same as that of the
specific heat coefficient, only with a different scaling function:
\be
N(t,T) = t^{2\nu - 1}\,f_N(T/t)\ .
\label{eq:8}
\ee
This behavior of the density of states, if it exists, is measureable by means of tunneling. 
For all $\nu > 1/2$ Eq.\ (\ref{eq:8}) predicts a vanishing density of states at the Fermi
surface, or a pseudogap \cite{Mott_1968}. Note that this is consistent with the notion of a diverging
effective mass $m^*$: Consider the standard quasi-particle picture of Landau Fermi-liquid
theory \cite{Abrikosov_Gorkov_Dzyaloshinski_1963}. Let $m$ be the bare electron mass,
$m^*$ the renormalized, or effective, mass, and $Z$ the quasiparticle weight. For simplicity,
consider a wave number-independent self energy. Then $m^*/m = 1/Z$, and thus $Z\to 0$,
which implies a vanishing density of states, corresponds to a diverging $m^*$. (This is not
the only mechanism that can lead to a diverging $m^*$ \cite{Asgari_et_al_2005,
Zhang_DasSarma_2005}, but it is one possibility.) The electrical resistivity also shows
critical behavior at this transition \cite{Kirkpatrick_Belitz_2012}, but presumably
this would be masked by non-critical scattering mechanisms as discussed above.

Finally, we come back to the point that there are multiple dynamical
exponents, one of which is $z_c=1$, which describes the density 
dynamics \cite{multiple_z_footnote}.  If this is the dominant dynamical exponent
for the observables discussed above, then $\nu = y_h = z = 1$, and both the
specific heat coefficient and the density of states at $T=0$ will vanish linearly
as a function of $t$ \cite{DIV_footnote}.

We now summarize and discuss our results. Starting from the experimental
observation that the thermopower displays critical behavior in a $2$-$d$ electron
system, we have employed simple scaling arguments to make predictions about
other observables that allow for confirming or refuting the notion that the
experiment of Ref.\ \onlinecite{Mokashi_et_al_2012} is indeed indicative of
a quantum phase transition that is caused by strong correlations in the electron
fluid, and insensitive to disorder. Scaling shows that the thermopower experiment
yields the product of the correlation length exponent $\nu$ and the dynamical 
exponent $z$. It predicts that the specific heat coefficient will also display
scaling behavior, which will allow for a separate determination of $\nu$. If
$\nu$ were found to violate the Harris criterion, this would definitively rule out a 
disorder-dominated nature of the transition. The unknown order parameter
density is predicted to obey scaling characterized by a third independent exponent 
$y_h$. These predictions are all very general, and hinge only on the assumption
that the observed behavior of the thermopower does indeed reflect a true
quantum phase transition. In addition, we have given a criterion to check whether this 
transition is a manifestation of the Fermi-liquid-to-non-Fermi-liquid transition
discussed in Ref.\ \onlinecite{Kirkpatrick_Belitz_2012}. In that case, the density
of states at the Fermi level is predicted to be critical, and its scaling behavior will be
the same as that of the specific heat coefficient, except for a different functional
form of the scaling function. In particular, at $T=0$ both quantities will
vanish as $(n - n_{\text{c}})^{2\nu - 1}$, with $\nu$ the correlation length
exponent, as the transition is approached from high densities. 

This work was supported by the National Science Foundation under Grant Nos.
DMR-09-29966 and DMR-09-01907.

%\bibliography{thermopower}

\begin{thebibliography}{33}
\expandafter\ifx\csname natexlab\endcsname\relax\def\natexlab#1{#1}\fi
\expandafter\ifx\csname bibnamefont\endcsname\relax
  \def\bibnamefont#1{#1}\fi
\expandafter\ifx\csname bibfnamefont\endcsname\relax
  \def\bibfnamefont#1{#1}\fi
\expandafter\ifx\csname citenamefont\endcsname\relax
  \def\citenamefont#1{#1}\fi
\expandafter\ifx\csname url\endcsname\relax
  \def\url#1{\texttt{#1}}\fi
\expandafter\ifx\csname urlprefix\endcsname\relax\def\urlprefix{URL }\fi
\providecommand{\bibinfo}[2]{#2}
\providecommand{\eprint}[2][]{\url{#2}}

\bibitem[{\citenamefont{Abrahams et~al.}(1979)\citenamefont{Abrahams, Anderson,
  Licciardello, and Ramakrishnan}}]{Abrahams_et_al_1979}
\bibinfo{author}{\bibfnamefont{E.}~\bibnamefont{Abrahams}},
  \bibinfo{author}{\bibfnamefont{P.~W.} \bibnamefont{Anderson}},
  \bibinfo{author}{\bibfnamefont{D.~C.} \bibnamefont{Licciardello}},
  \bibnamefont{and} \bibinfo{author}{\bibfnamefont{T.~V.}
  \bibnamefont{Ramakrishnan}}, \bibinfo{journal}{Phys. Rev. Lett.}
  \textbf{\bibinfo{volume}{42}}, \bibinfo{pages}{673} (\bibinfo{year}{1979}).

\bibitem[{\citenamefont{Wegner}(1979)}]{Wegner_1979}
\bibinfo{author}{\bibfnamefont{F.}~\bibnamefont{Wegner}}, \bibinfo{journal}{Z.
  Phys. B} \textbf{\bibinfo{volume}{35}}, \bibinfo{pages}{207}
  (\bibinfo{year}{1979}).

\bibitem[{AT_()}]{AT_footnote}
\bibinfo{note}{It was later found that there is one universality class of
  Anderson transitions, realized in systems with strong spin-orbit scattering,
  that displays a normal metal phase in $d=2$ \cite{Evers_Mirlin_2008}.
  However, this is not believed to be relevant for the systems under
  consideration here, see Ref.\ \onlinecite{Abrahams_Kravchenko_Sarachik_2001}.
  We also note that by ``metal-insulator transition'' we mean a transition
  involving a normal metal phase, as opposed to the well-established
  superconductor-insulator transition in thin metal films \cite{Goldman_2010},
  and that the Quantum Hall Effect is caused by delocalized surface states and
  does not constitute a bulk phase transition \cite{Halperin_1982}.}

\bibitem[{\citenamefont{Finkelstein}(1984)}]{Finkelstein_1984a}
\bibinfo{author}{\bibfnamefont{A.~M.} \bibnamefont{Finkelstein}},
  \bibinfo{journal}{Zh. Eksp. Teor. Fiz.} \textbf{\bibinfo{volume}{86}},
  \bibinfo{pages}{367} (\bibinfo{year}{1984}), \bibinfo{note}{[Sov. Phys. JETP
  {\bf 59}, 212 (1984)]}.

\bibitem[{\citenamefont{Belitz and
  Kirkpatrick}(1994)}]{Belitz_Kirkpatrick_1994}
\bibinfo{author}{\bibfnamefont{D.}~\bibnamefont{Belitz}} \bibnamefont{and}
  \bibinfo{author}{\bibfnamefont{T.~R.} \bibnamefont{Kirkpatrick}},
  \bibinfo{journal}{Rev. Mod. Phys.} \textbf{\bibinfo{volume}{66}},
  \bibinfo{pages}{261} (\bibinfo{year}{1994}).

\bibitem[{\citenamefont{Kravchenko et~al.}(1994)\citenamefont{Kravchenko,
  Kravchenko, Furneaux, Pudalov, and M.D'Iorio}}]{Kravchenko_et_al_1994}
\bibinfo{author}{\bibfnamefont{S.~V.} \bibnamefont{Kravchenko}},
  \bibinfo{author}{\bibfnamefont{G.~V.} \bibnamefont{Kravchenko}},
  \bibinfo{author}{\bibfnamefont{J.~E.} \bibnamefont{Furneaux}},
  \bibinfo{author}{\bibfnamefont{V.~M.} \bibnamefont{Pudalov}},
  \bibnamefont{and} \bibinfo{author}{\bibnamefont{M.D'Iorio}},
  \bibinfo{journal}{Phys. Rev. B} \textbf{\bibinfo{volume}{50}},
  \bibinfo{pages}{8039} (\bibinfo{year}{1994}).

\bibitem[{\citenamefont{Abrahams et~al.}(2001)\citenamefont{Abrahams,
  Kravchenko, and Sarachik}}]{Abrahams_Kravchenko_Sarachik_2001}
\bibinfo{author}{\bibfnamefont{E.}~\bibnamefont{Abrahams}},
  \bibinfo{author}{\bibfnamefont{S.~V.} \bibnamefont{Kravchenko}},
  \bibnamefont{and} \bibinfo{author}{\bibfnamefont{M.~P.}
  \bibnamefont{Sarachik}}, \bibinfo{journal}{Rev. Mod. Phys.}
  \textbf{\bibinfo{volume}{73}}, \bibinfo{pages}{251} (\bibinfo{year}{2001}).

\bibitem[{\citenamefont{Spivak et~al.}(2010)\citenamefont{Spivak, Kravchenko,
  Kivelson, and Gao}}]{Spivak_et_al_2010}
\bibinfo{author}{\bibfnamefont{B.}~\bibnamefont{Spivak}},
  \bibinfo{author}{\bibfnamefont{S.~V.} \bibnamefont{Kravchenko}},
  \bibinfo{author}{\bibfnamefont{S.~A.} \bibnamefont{Kivelson}},
  \bibnamefont{and} \bibinfo{author}{\bibfnamefont{X.~P.~A.}
  \bibnamefont{Gao}}, \bibinfo{journal}{Rev. Mod. Phys.}
  \textbf{\bibinfo{volume}{82}}, \bibinfo{pages}{1743} (\bibinfo{year}{2010}).

\bibitem[{\citenamefont{Mokashi et~al.}(2012)\citenamefont{Mokashi, Li, Wen,
  Kravchenko, Shashkin, Dolgopolov, and Sarachik}}]{Mokashi_et_al_2012}
\bibinfo{author}{\bibfnamefont{A.}~\bibnamefont{Mokashi}},
  \bibinfo{author}{\bibfnamefont{S.}~\bibnamefont{Li}},
  \bibinfo{author}{\bibfnamefont{B.}~\bibnamefont{Wen}},
  \bibinfo{author}{\bibfnamefont{S.}~\bibnamefont{Kravchenko}},
  \bibinfo{author}{\bibfnamefont{A.~A.} \bibnamefont{Shashkin}},
  \bibinfo{author}{\bibfnamefont{V.~T.} \bibnamefont{Dolgopolov}},
  \bibnamefont{and} \bibinfo{author}{\bibfnamefont{M.~P.}
  \bibnamefont{Sarachik}}, \bibinfo{journal}{Phys. Rev. Lett.}
  \textbf{\bibinfo{volume}{109}}, \bibinfo{pages}{096405}
  (\bibinfo{year}{2012}).

\bibitem[{Kub()}]{Kubo_formula_footnote}
\bibinfo{note}{In a technical description in terms of Kubo functions, the
  thermopower is proportional to a heat current -- number current correlation
  function divided by a number current -- number current correlation function
  times the temperature \cite{Mahan_1981}. While it is not necessary to know
  this for the simple power-counting arguments employed in this paper, it is
  consistent with them.}

\bibitem[{\citenamefont{Abanov and Chubukov}(2004)}]{Abanov_Chubukov_2004}
\bibinfo{author}{\bibfnamefont{A.}~\bibnamefont{Abanov}} \bibnamefont{and}
  \bibinfo{author}{\bibfnamefont{A.~V.} \bibnamefont{Chubukov}},
  \bibinfo{journal}{Phys. Rev. Lett.} \textbf{\bibinfo{volume}{93}},
  \bibinfo{pages}{255702} (\bibinfo{year}{2004}).

\bibitem[{\citenamefont{Kivelson et~al.}(2003)\citenamefont{Kivelson, Bindloss,
  Fradkin, Oganesyan, Tranquada, Kapitulnik, and Howald}}]{Kivelson_et_al_2003}
\bibinfo{author}{\bibfnamefont{S.~A.} \bibnamefont{Kivelson}},
  \bibinfo{author}{\bibfnamefont{I.~P.} \bibnamefont{Bindloss}},
  \bibinfo{author}{\bibfnamefont{E.}~\bibnamefont{Fradkin}},
  \bibinfo{author}{\bibfnamefont{V.}~\bibnamefont{Oganesyan}},
  \bibinfo{author}{\bibfnamefont{J.~M.} \bibnamefont{Tranquada}},
  \bibinfo{author}{\bibfnamefont{A.}~\bibnamefont{Kapitulnik}},
  \bibnamefont{and} \bibinfo{author}{\bibfnamefont{C.}~\bibnamefont{Howald}},
  \bibinfo{journal}{Rev. Mod. Phys.} \textbf{\bibinfo{volume}{75}},
  \bibinfo{pages}{1201} (\bibinfo{year}{2003}).

\bibitem[{\citenamefont{Stanley}(1971)}]{Stanley_1971}
\bibinfo{author}{\bibfnamefont{E.}~\bibnamefont{Stanley}},
  \emph{\bibinfo{title}{Introduction to Phase Transitions and Critical
  Phenomena}} (\bibinfo{publisher}{Oxford University Press, Oxford},
  \bibinfo{year}{1971}).

\bibitem[{\citenamefont{Ma}(1976)}]{Ma_1976}
\bibinfo{author}{\bibfnamefont{S.-K.} \bibnamefont{Ma}},
  \emph{\bibinfo{title}{Modern Theory of Critical Phenomena}}
  (\bibinfo{publisher}{Benjamin, Reading, MA}, \bibinfo{year}{1976}).

\bibitem[{\citenamefont{Kirkpatrick and
  Belitz}(2012)}]{Kirkpatrick_Belitz_2012}
\bibinfo{author}{\bibfnamefont{T.~R.} \bibnamefont{Kirkpatrick}}
  \bibnamefont{and} \bibinfo{author}{\bibfnamefont{D.}~\bibnamefont{Belitz}},
  \bibinfo{journal}{Phys. Rev. Lett.} \textbf{\bibinfo{volume}{108}},
  \bibinfo{pages}{086404} (\bibinfo{year}{2012}).

\bibitem[{hid()}]{hidden_order_footnote}
\bibinfo{note}{Historically, there often have been indications of a phase
  transition before the order parameter was identified. A famous historical
  example is antiferromagnetism, a more recent one, the ``hidden order'' in the
  heavy-fermion metal URu$_2$Si$_2$, see Ref.
  \onlinecite{Mydosh_Oppeneer_2011}.}

\bibitem[{\citenamefont{Sachdev}(1999)}]{Sachdev_1999}
\bibinfo{author}{\bibfnamefont{S.}~\bibnamefont{Sachdev}},
  \emph{\bibinfo{title}{Quantum Phase Transitions}}
  (\bibinfo{publisher}{Cambridge University Press, Cambridge},
  \bibinfo{year}{1999}).

\bibitem[{ine()}]{inequality_footnote}
\bibinfo{note}{We also note the inequality $e\vert S\vert \leq
  \sqrt{\kappa/T\sigma}$, with $\sigma$ and $\kappa$ the electrical and thermal
  conductivities, respectively, which follows from the requirement that the
  entropy production rate must be positive, see Ref.\ \onlinecite{Onsager_1931}
  and the reference to Boltzmann therein. (Here $\kappa$ and $\sigma$ are
  defined strictly analogously, viz, as the transport coefficients for
  vanishing electrochemical potential and temperature gradients, respectively.
  Different definitions for $\kappa$ can occasionally be found in the
  literature.) Since $\kappa/T$ and $\sigma$ have the same scale dimension by
  power counting, this relation is consistent with the notion that $eS$ has a
  vanishing scale dimension.}

\bibitem[{mul()}]{multiple_z_footnote}
\bibinfo{note}{At this point we must stress that in general there is more than
  one dynamical exponent due to different time scales in the system
  \cite{Belitz_Kirkpatrick_1994}. For instance, the Coulomb interaction implies
  that there is a time scale for the charge or density dynamics that is
  characterized by $z_c = 1$, whereas the critical dynamical exponent is in
  general $z \neq 1$. Which $z$ enters in any given context cannot be
  determined from scaling considerations alone.}

\bibitem[{ene()}]{energy_scales_footnote}
\bibinfo{note}{For scaling plots of this sort one has to keep in mind that the
  dynamical critical scaling can be expected to hold only at temperatures small
  compared to the Fermi temperature, which sets the microscopic energy scale.
  Since the Fermi temperature in the samples in question is only on the order
  of a few Kelvin \cite{Abrahams_Kravchenko_Sarachik_2001}, the temperatures of
  a few hundred mK achieved in Ref.\ \onlinecite{Mokashi_et_al_2012} may not
  suffice to demonstrate the characteristic $T/t$ scaling behavior.}

\bibitem[{\citenamefont{Harris}(1974)}]{Harris_1974}
\bibinfo{author}{\bibfnamefont{A.~B.} \bibnamefont{Harris}},
  \bibinfo{journal}{J. Phys. C} \textbf{\bibinfo{volume}{7}},
  \bibinfo{pages}{1671} (\bibinfo{year}{1974}).

\bibitem[{\citenamefont{Chayes et~al.}(1986)\citenamefont{Chayes, Chayes,
  Fisher, and Spencer}}]{Chayes_et_al_1986}
\bibinfo{author}{\bibfnamefont{J.}~\bibnamefont{Chayes}},
  \bibinfo{author}{\bibfnamefont{L.}~\bibnamefont{Chayes}},
  \bibinfo{author}{\bibfnamefont{D.~S.} \bibnamefont{Fisher}},
  \bibnamefont{and} \bibinfo{author}{\bibfnamefont{T.}~\bibnamefont{Spencer}},
  \bibinfo{journal}{Phys. Rev. Lett.} \textbf{\bibinfo{volume}{57}},
  \bibinfo{pages}{2999} (\bibinfo{year}{1986}).

\bibitem[{\citenamefont{Mott}(1968)}]{Mott_1968}
\bibinfo{author}{\bibfnamefont{N.~F.} \bibnamefont{Mott}},
  \bibinfo{journal}{Rev. Mod. Phys.} \textbf{\bibinfo{volume}{40}},
  \bibinfo{pages}{677} (\bibinfo{year}{1968}).

\bibitem[{\citenamefont{Abrikosov et~al.}(1963)\citenamefont{Abrikosov, Gorkov,
  and Dzyaloshinski}}]{Abrikosov_Gorkov_Dzyaloshinski_1963}
\bibinfo{author}{\bibfnamefont{A.~A.} \bibnamefont{Abrikosov}},
  \bibinfo{author}{\bibfnamefont{L.~P.} \bibnamefont{Gorkov}},
  \bibnamefont{and} \bibinfo{author}{\bibfnamefont{I.~E.}
  \bibnamefont{Dzyaloshinski}}, \emph{\bibinfo{title}{Methods of Quantum Field
  Theory in Statistical Physics}} (\bibinfo{publisher}{Dover, New York},
  \bibinfo{year}{1963}).

\bibitem[{\citenamefont{Asgari et~al.}(2005)\citenamefont{Asgari, Davoudi,
  Polini, Guiliani, Tosi, and Vignale}}]{Asgari_et_al_2005}
\bibinfo{author}{\bibfnamefont{R.}~\bibnamefont{Asgari}},
  \bibinfo{author}{\bibfnamefont{B.}~\bibnamefont{Davoudi}},
  \bibinfo{author}{\bibfnamefont{M.}~\bibnamefont{Polini}},
  \bibinfo{author}{\bibfnamefont{G.~F.} \bibnamefont{Guiliani}},
  \bibinfo{author}{\bibfnamefont{M.~P.} \bibnamefont{Tosi}}, \bibnamefont{and}
  \bibinfo{author}{\bibfnamefont{G.}~\bibnamefont{Vignale}},
  \bibinfo{journal}{Phys. Rev. B} \textbf{\bibinfo{volume}{71}},
  \bibinfo{pages}{045323} (\bibinfo{year}{2005}).

\bibitem[{\citenamefont{Zhang and {Das~Sarma}}(2005)}]{Zhang_DasSarma_2005}
\bibinfo{author}{\bibfnamefont{Y.}~\bibnamefont{Zhang}} \bibnamefont{and}
  \bibinfo{author}{\bibfnamefont{S.}~\bibnamefont{{Das~Sarma}}},
  \bibinfo{journal}{Phys. Rev. B} \textbf{\bibinfo{volume}{71}},
  \bibinfo{pages}{045322} (\bibinfo{year}{2005}).

\bibitem[{DIV()}]{DIV_footnote}
\bibinfo{note}{We stress again that scaling considerations alone cannot
  determine the value of $z$ in any given context; $z=1$ represents an
  additional assumption. By contrast, the homogeneity laws expressed by Eqs.\
  (\ref{eqs:3}), (\ref{eq:5}), (\ref{eq:6}), and, to the extent that the
  density of states shows critical behavior, (\ref{eq:8}), are expected to be
  generally valid unless simple scaling is invalidated, e.g., by the presence
  of dangerous irrelevant variables.}

\bibitem[{\citenamefont{Evers and Mirlin}(2008)}]{Evers_Mirlin_2008}
\bibinfo{author}{\bibfnamefont{F.}~\bibnamefont{Evers}} \bibnamefont{and}
  \bibinfo{author}{\bibfnamefont{A.~D.} \bibnamefont{Mirlin}},
  \bibinfo{journal}{Rev. Mod. Phys.} \textbf{\bibinfo{volume}{80}},
  \bibinfo{pages}{1355} (\bibinfo{year}{2008}).

\bibitem[{\citenamefont{Goldman}(2010)}]{Goldman_2010}
\bibinfo{author}{\bibfnamefont{A.~M.} \bibnamefont{Goldman}},
  \bibinfo{journal}{Int. J. Mod. Phys. B} \textbf{\bibinfo{volume}{24}},
  \bibinfo{pages}{4081} (\bibinfo{year}{2010}).

\bibitem[{\citenamefont{Halperin}(1982)}]{Halperin_1982}
\bibinfo{author}{\bibfnamefont{B.~I.} \bibnamefont{Halperin}},
  \bibinfo{journal}{Phys. Rev. B} \textbf{\bibinfo{volume}{25}},
  \bibinfo{pages}{2185} (\bibinfo{year}{1982}).

\bibitem[{\citenamefont{Mahan}(1981)}]{Mahan_1981}
\bibinfo{author}{\bibfnamefont{G.~D.} \bibnamefont{Mahan}},
  \emph{\bibinfo{title}{Many-Particle Physics}} (\bibinfo{publisher}{Plenum,
  New York}, \bibinfo{year}{1981}).

\bibitem[{\citenamefont{Mydosh and Oppeneer}(2011)}]{Mydosh_Oppeneer_2011}
\bibinfo{author}{\bibfnamefont{J.}~\bibnamefont{Mydosh}} \bibnamefont{and}
  \bibinfo{author}{\bibfnamefont{P.~M.} \bibnamefont{Oppeneer}},
  \bibinfo{journal}{Rev. Mod. Phys.} \textbf{\bibinfo{volume}{83}},
  \bibinfo{pages}{1301} (\bibinfo{year}{2011}).

\bibitem[{\citenamefont{Onsager}(1931)}]{Onsager_1931}
\bibinfo{author}{\bibfnamefont{L.}~\bibnamefont{Onsager}},
  \bibinfo{journal}{Phys. Rev.} \textbf{\bibinfo{volume}{37}},
  \bibinfo{pages}{405} (\bibinfo{year}{1931}).

\end{thebibliography}

\end{document}